\documentstyle[psfig,12pt,aaspp4]{article}

\def\la{\hbox{\rlap{\raise.3ex\hbox{$<$}}\lower.8ex\hbox{$\sim$}\ }}
\def\ga{\hbox{\rlap{\raise.3ex\hbox{$>$}}\lower.8ex\hbox{$\sim$}\ }}

\def\pb{\parbox}
\def\pb5{\parbox[t]{4.5cm}}

\def\cm2sec{\ cm$^{2}$-sec}

\def\H2S1{H$_2$ $S$(1)}

\def\apjl{{\it Ap.J. Letters}}

\def\apjs{{\it Ap.\ J.\ Suppl.}}

\begin{document}

\title{\bf Simulations and Measurements of the Background Encountered by a
High-Altitude Balloon-Borne Experiment for Hard X-ray Astronomy}

\author{\bf Kenneth S. K. Lum, Joseph J. Mohr\footnote{Postal address: Univ. of Michigan, Dept. of Physics, 500 E. University, Ann Arbor, MI 48109, USA}, Didier Barret\footnote{Postal address: Centre d'Etude Spatiale des Rayonnements, CESR-CNRS/UPS, 9 Av. du colonel Roche, BP 4346, Toulouse, Cedex, France},
Jonathan E. Grindlay, and Raj P. Manandhar\footnote{Postal address: NASA/Marshall Space Flight Center, Space Science Laboratory, Huntsville, AL 35812, USA}}

\affil{Harvard-Smithsonian Center for Astrophysics, 60 Garden St., Cambridge, MA 02138, USA}

\author{Submitted to\\{\it Nuclear Instruments and Methods in Physics
Research, Section A}}

\date{\today}

\begin{abstract}

We have modelled the hard X-ray background expected for a high-altitude
balloon flight of the Energetic X-ray Telescope Experiment (EXITE2),
an imaging phoswich detector/telescope for the 20--600 keV energy range.
Photon
and neutron-induced contributions to the background are considered.
We describe the code and the results
of a series of simulations with different shielding configurations.  The
simulated hard X-ray background for the actual flight configuration agrees
reasonably well (within a factor of $\sim$ 2) with the results measured on the
first flight of EXITE2 from Palestine, Texas.  The measured
background flux at 100 keV
is $\sim$ 4 $\times$ 10$^{-4}$ counts cm$^{-2}$ s$^{-1}$ keV$^{-1}$.\\
\newline
\newline
\newline
\noindent
Send proofs to:\\
\noindent
Kenneth S. K. Lum\newline
\noindent
Harvard-Smithsonian Center for Astrophysics\newline
\noindent
60 Garden St., MS83\newline
\noindent
Cambridge, MA 02138, USA\newline
\noindent
Tel: 617-267-6645\newline
\noindent
Fax: 617-267-8284\newline
\noindent
Email: {\it ksl@head-cfa.harvard.edu}\\

\noindent
Keywords: Detectors, astrophysics, X-ray astronomy, gamma-ray astronomy

\end{abstract}

\newpage
\section{Introduction}

Hard X-ray ($\sim$ 20-600 keV) detectors for space-based astronomy
encounter background arising from various
forms of ionizing radiation and their interactions with the surrounding
payload and the detector itself.  Because detection sensitivities are
strongly dependent on this background, it must be carefully considered in the
design of these instruments and interpretation of data.
As part of the process of optimizing the design of the second
Energetic X-ray Imaging Telescope Experiment (EXITE2) [1, 2],
we developed a Monte-Carlo simulation code which
calculates the expected in-flight background.
The program propagates incident photons through the payload,
determines the spectral and spatial distributions of photons which enter the
detector, incorporates the detector response, and produces the energy spectrum
of detected photon-induced background events.  A separate code estimates the
neutron-induced background.  Because a variety of detector and shielding
configurations were to be considered for optimization, we designed the program
so that the
spatial and energy characteristics of the incident photon spectrum and the
materials and geometry of the payload are all specified in straightforward
input parameter files which can be easily modified.
The background simulation does not include charged-particle events,
which are efficiently rejected by either the upper-level discriminator
or by the active shields, as demonstrated by previous in-flight
measurements made by the EXITE1 detector [3, 4].

Here we summarize the EXITE2 payload ($\S2$), describe the Monte-Carlo
simulation ($\S3$), detail predicted backgrounds for a variety of payload
configurations ($\S4$), and then compare the predicted background, for
the appropriate configuration, to data ($\S5$) acquired during the June 1993
engineering test flight of EXITE2 from Palestine, Texas.

\section{EXITE2 Payload}

A schematic of the EXITE2 payload in the flight configuration is shown in
Figure 1.  Full details of the original EXITE2 detector configuration, as used
to obtain the flight background data discussed here ($\S5$), are given in [2],
and only a brief summary is mentioned here.
The detector itself is located between the collimator and the
photomultiplier tube (PMT) array and consists of a 36 cm $\times$ 36 cm
$\times$ 1 cm thick NaI(Tl) crystal surrounded on five sides by 2 cm of
CsI(Na), with a 0.25 mm thick aluminum entrance window and a 1.905 cm thick
glass exit window.  Thus, the detector as a whole is 40 cm $\times$ 40 cm
$\times$ 4.93 cm.  The detector and front-end electronics are enclosed in
a sealed vessel pressurized to slightly above 1 atm, with a 0.35 mm thick
Mylar entrance window, which is pressure-supported by the overlying
collimator assembly.

The collimator consists of 52 interlocking blades (26 aligned in each
orthogonal direction) spaced 1.6 cm apart (center-to-center) and each a
laminate composed of a 0.9 mm layer of lead sandwiched between two 0.2 mm
layers of 260 brass (70\% copper, 30\% zinc).  The collimator field of view is
4.7$^{\circ}$ (FWHM).  The mask pixels are each 1.6 cm $\times$ 1.6 cm,
commensurate with the collimator cells, and are a graded stack with 12 mm lead,
1 mm tin, and 1 mm copper.
The mask is supported by two pieces of 1.5 mm thick Lexan
which sandwich the mask pixels.
With a mask-detector distance or focal length of 2.5 m, the telescope has an angular
resolution of 22$^{\prime}$.

The side and rear passive shields consist of 3.175 mm lead, 2 mm tin, and 1
mm copper.  Active shielding composed of 1.588 cm thick NE102 plastic
scintillator surrounds most of the sides and rear of the detector and is used
to reject charged-particle background.  The active shields are each read out
by a single two-inch diameter PMT (Hamamatsu R1306) optically coupled
through a light guide and edge readout.

The main detector readout consists of a 7 $\times$ 7 array of PMTs.  The pulse
rise
time of each event is measured and used to distinguish between events absorbed
in the NaI(Tl) and those absorbed in the CsI(Na).  The detector is continuously
calibrated using a system of twelve light-emitting diodes (LEDs) whose pulsed
($\sim$ 10 Hz each) light
enters the detector from the rear through optical fibers, four $^{241}$Am
sources which are built into the bottom of the collimator near the four corners
of the detector, and periodic ($\sim$ 130 Hz) digitizations of the
individual-channel
baselines (``CAL events'') to monitor the offsets and overall detector
baseline.  In addition, when the telescope
is in the stowed position, the detector is partially illuminated by a
$^{109}$Cd source
mounted above, on the gondola.

\section{Simulating the Detector Background}

The sources of detected background in balloon-borne hard X-ray detectors are
reviewed in detail by
Geh\-rels [5] and Dean {\it et al.} [6].  We include the effects of
atmospheric and cosmic gamma rays, and atmospheric and locally produced
neutrons.  Details are given in the following subsections.

\subsection{Photon-Induced Background}

The photon-induced background is modelled by a two-part simulation code.
The first part of the code consists of a photon-transport
program, {\it phot\_trans} (developed initially at the Massachusetts Institute
of Technology and then by us at Harvard), which takes input spectral and spatial
distributions of gamma rays and from these generates the spectral and spatial
distributions of photons incident on the actual detector.  For the input
spectrum, we use the one given by Gehrels [5], which is based on data
obtained at balloon altitudes ($\sim$ 3.5 g cm$^{-2}$ residual atmosphere) over
Palestine, Texas.  The incident photons are assumed to be isotropic, which
is a good approximation [5].  A
simplified flow diagram of {\it phot\_trans} is shown in Figure 2.

The {\it phot\_trans} code takes into account K-shell fluorescence following
photoelectric absorption.  It includes $K_{\alpha}$ and $K_{\beta}$ lines,
which dominate, and neglects higher-order K lines as well as lower energy
(e.g., L shell) transitions.  The total fluorescence
yields assumed are those given by Krause [7].  The relative intensities of
the individual fluorescence lines are those given by Kaye and Laby [8].
The L and higher-order fluorescence lines have energies which are much lower
than the K-line energies (e.g., $\sim$ 10--15 keV vs. $\sim$ 73--87 keV, for
lead) and outside the energy range of interest for our experiment.  We
therefore neglect L and higher-order fluorescence.  The Auger effect is
neglected because Auger electrons rapidly lose their energy in traversing the
material of origin and do not cause significant X-ray emission.
{\it phot\_trans} uses Compton-scattering cross sections derived from the
Klein-Nishina formula for unpolarized radiation [9].  The code is fully
three-dimensional and allows for any materials and configuration (e.g.,
detector frame and housing, side shields) around the detector.

The photons which arrive at the detector are then detected by a simulation
code, based on EGS4 [10, 11],
which determines the positions and amounts of energy deposition in the
detector.  The known detector response, which has been measured in the
laboratory at several energies using collimated (to $\sim$ 2 mm on the
detector) $^{241}$Am and $^{133}$Ba sources placed
at 25 $\times$ 25 locations on a grid of points 1.6 cm apart, is then used to
produce the predicted photon contribution to the detected energy spectrum.

\subsection{Neutron-Induced Background}

The neutron-induced background is estimated using the semi-empirical method
described by Dean {\it et al.} [6], which uses data from the MISO NaI(Tl)
gamma-ray telescope [12].  The method first calculates the locally produced
neutron flux incident on the detector resulting from cosmic-ray interactions
with each part of the surrounding payload.  Cosmic rays interact with
nuclei in the payload materials, generating free neutrons which are emitted
isotropically, some of which enter the detector.
An atmospheric contribution to the neutron flux,
based on the calculations by Armstrong {\it et al.} (1973) [13],
is added to the local contribution.
The induced detector background spectrum per neutron (in
counts cm$^{-3}$ s$^{-1}$ keV$^{-1}$ neutron$^{-1}$) was determined from
a background measurement made by the MISO instrument, which was
flown on a stratospheric balloon flight (3.5 g cm$^{-2}$)
over Palestine, Texas.
%Because the MISO telescope was designed with a massive shield, its background
%was dominated by locally produced neutrons.
The neutron-induced background spectrum for other NaI(Tl) and CsI(Na) detectors
can be estimated from the MISO results by taking into account the mass
distribution of the corresponding payload and the geometry of the detector and
rescaling the entire spectrum, which is approximately independent of energy
below $\sim$ 600 keV.  We have used
this method to estimate the neutron-induced background in the EXITE2 detector.
%By this method, the spectral
%shape of the neutron-induced background is determined primarily by
%the MISO measurement and to a lesser extent by the mass distribution of the
%EXITE2 payload, while the normalization is determined by the geometry of the
%EXITE2 detector.

\section{Background Predictions}

The code for the photon-induced background was run for a number of possible
EXITE2 shielding configurations.  In particular, we investigated the effects of
varying the thicknesses of layers in the mask, collimator, and passive shields,
and then chose a mechanical design which yielded acceptable background
characteristics while minimizing weight.
The simulations included the part of the payload immediately surrounding the
detector but none of the surrounding gondola structure or the MIXE1 detector
[14] flown adjacent to the EXITE2 detector.

The results of twelve of
the simulations, each with 10$^{5}$ photons incident on the detector in the
0.010--2 MeV energy range, are summarized in Tables 1 and 2, and Figures
3(a--d) and 4(a--d), where the area of the detector is taken to be 1600
cm$^{2}$, the geometrical area of the entire crystal assembly as viewed from
the front of the detector.  The results shown in these tables and figures
are for photons {\it incident} on the detector.  They do not include the
detector response and are without phoswich rejection of CsI(Na) events, and
thus the
results given here should be more directly applicable to considerations for
other payloads.  The results shown in
Table 1 and Figures 3(a--d) are for photons incident on {\it all} external
surfaces of the detector (i.e., for all photons entering the detector), in
order to characterize the total background incident on the detector.
The results shown in Table 2 are for photons
incident only on the front (cases 1--6), side (cases 7--9), and rear (cases
10--12) surfaces of the detector; from these results, it should be easier to
infer the effects of individual payload shielding components.  The results
shown in
Figures 4(a--d) are for photons incident only on the front (4[a--b]), side
(4[c]), and rear (4[d]) surfaces of the detector.
The binning in Figures 4(a--d) is coarser than in Figures 3(a--d)
in order to improve statistics.
Each ``case'' listed in the tables represents the same payload
configuration in both tables.  For example, case 1 represents the flight
configuration in both tables, and case 5 represents the configuration with the
collimator removed in both tables.
Case 1 was done with the EXITE2
payload in the flight configuration.  We varied the construction of the mask,
collimator, side shields, and rear shield in cases 2--4, 5--6, 7--9, and
10--12, respectively.  Although
the actual collimator was made with brass sandwiching the lead, in the
simulations we replaced the brass with copper, which has similar X-ray
absorption properties and density but is a single element and thus easier to
implement.

Each of the spectra shown in Figures 3(a--d) and 4(a--d) is made up of
interacting photons
which are a product of interactions in the surrounding payload,
and non-interacting photons which have passed through the
surrounding payload or collimator aperture without interacting with any of the
material before entering the detector.  In Figures 5(a--c), we show the spectra
of interacting and non-interacting photons, and the sum of the two, for photons
incident on the front (a), side (b), and rear (c) surfaces of the detector,
based on a simulation with 10$^{6}$ photons incident on the detector from
0.010--2 MeV, with the payload in its flight configuration.

\section{Background Measurements}

EXITE2 was flown on an engineering test flight on 13--14 June 1993
from Palestine, Texas.  Background data were accumulated while the payload was
ascending to float altitude ($\sim$ 3.8 g cm$^{-2}$) and for several hours
afterwards.  These data have been analyzed and
used to produce background spectra which were compared to the simulated
results.

The procedure used to analyze the data was as follows.  The data were divided
into subintervals each corresponding to $\sim$ 10 minutes of data.  For each
subinterval, the
$^{241}$Am, LED, and CAL-event data were used to determine the gains
and offsets of the 49 PMT channels [15].
The individual-channel pulse heights were
thus corrected for electronic gain and offset variations, which were
small---typically $\sim$ 2\% and $\sim$ 200 eV (i.e., 0.2\% of the PMT signal
amplitude
in the case where a 100 keV X-ray is absorbed in the detector directly above
a given PMT), respectively.  Pulse-shape
discrimination (PSD) was used to reject events which were absorbed in the
CsI(Na).  The PSD rejection efficiency (i.e., percentage of CsI[Na] events
rejected while retaining at least 90\% of the NaI[Tl] events) ranges from
$\sim$ 30\% at 20 keV to $>$ 99\% at energies $>$ $\sim$
60 keV and is a function of how well the NaI(Tl) and CsI(Na) pulse shapes
can be distinguished from each other at different signal amplitudes.
The number of detected LED events was compared to the number known to have
occurred during each subinterval (at $\sim$ 120 Hz) in order to
determine the electronic dead time of the system, for which a correction was
made.  The uncertainty in the measured background flux due to this correction
is $\sim$ 15\%.  The pulse height for each event was converted into an
energy, with a correction made for the position-dependent gain of the detector
as well as the non-linear response of NaI(Tl) for its optical output as a
function of energy [16].

Three background spectra were produced from the flight data and are displayed
in Figures 6(a--c) along with the simulated background spectrum corresponding
to the actual payload configuration.  Since in our simulations we did not take
into account the surrounding gondola structure (i.e., only the detector
frame and shielding are included), the same simulated
results are shown in all three figures.
The data used to produce the background
spectra in Figures 6(a--b) were acquired between 19:30 and 20:10 UT on 13
June 1993, when the telescope was approaching float ($\sim$ 5--10 g cm$^{-2}$)
but still in the stowed (i.e., vertical)
position.  Position cuts were used to remove $^{241}$Am source counts (from
the four calibration sources) in Figure
6(b).  The data used to produce the background spectrum in Figure 6(c) were
acquired between 20:18 and 20:54 UT, when the telescope had reached float
altitude ($\sim$ 3.8 g cm$^{-2}$) and was unstowed at an
elevation of $\sim$ 60$^{\circ}$, and the $^{109}$Cd source was outside
the field of view.  Position cuts were again used to remove
$^{241}$Am source counts.
The fluxes shown in Figures 6(a--c) correspond to an area of 1200 cm$^2$,
which takes into account the exclusion of $\sim$ 7\% of the NaI(Tl) area,
where the $^{241}$Am source counts lie.
The simulated spectrum includes
contributions from photon and neutron-induced events.
The simulated photon-induced background shown in these figures is based on a
simulation with 10$^6$ photons incident on the detector in the 0.010-2 MeV
energy range.

\section{Discussion}

A number of conclusions can be drawn from the results shown in Figures 3(a--d)
and 4(a--d).
The presence of the coded-aperture mask decreases the photon-induced
background.  The effect becomes less pronounced at higher energies, where a
larger fraction of the photons reaching the detector come from outside the
collimator field of view and thus do not pass through the mask.
The presence of the collimator reduces the background at
all energies, with greater effectiveness at lower energies, where only photons
passing through the collimator aperture can reach the detector.
Covering the sides of the lead in the collimator with copper
reduces the background slightly at all energies, but some lead-fluorescence
from the collimator blades persists.  Most of this is from the exposed ends of
the collimator blades, because the copper in the simulations, like the brass in
the actual collimator, does not extend over the ends of the blades, but, like
the lead, is flush with the detector pressure vessel entrance window; see
additional discussion below.
Figures 3(b) and 4(b) include an additional case, not listed in the
tables, in which the collimator consists only of lead and has had its pitch
changed from 1.6 cm to 5.5 cm so that the field of view is
$\sim$ 15$^{\circ}$.  The resulting background spectrum contains a much more
pronounced lead-fluorescence peak (19 $\sigma$ vs. 4 $\sigma$, in Figure 3[b]),
in reasonable
agreement with the results obtained by Schindler {\it et al.} [17] for the
Caltech detector with a similarly constructed lead collimator.  Our
results indicate that additional shielding of the collimator lead would be
warranted for instruments with wider fields of view.
The side and rear shields reduce the background
at all energies, as expected.  The results in Figure 4(d) illustrate the
importance of building a graded lead-tin-copper rear shield, where the tin
blocks lead fluorescence and the copper blocks tin fluorescence.

The spectra in Figures 5(a--c) show that at low energies ($\sim$ 20-40 keV) the
photon-induced background incident on the detector is completely dominated
by non-interacting photons which arrive at the front of the detector through
the collimator aperture.  The non-interacting component shown in Figure 5(a)
below $\sim$ 200 keV represents the ``aperture flux'' background which enters
through the collimator.  At higher
energies, the non-interacting component in Figure 5(a) is dominated by
photons which pass through collimator material, without interacting, and
reach the detector.  Figure 5(a) also shows the lead fluorescence lines at
75 keV and 85 keV which are from the lead in the collimator blades.  The
interacting components in Figures 5(b--c) represent the ``shield
leakage'' background and consist of incident and locally produced X-rays and
gamma-rays which leak through the shields and detector housing, and enter the
detector.

The lead-fluorescence fluxes shown in Table 2 (which include photons entering
the detector through the front [cases 1--6], side [cases 7--9], and rear
[cases 10--12] surfaces only) indicate that there would be
significant contributions from the mask (case 3), collimator
(case 6), side shields (case 8), and rear shield
(case 11) if the lead were not covered by tin.  Comparison of the 75 keV
fluxes shown for cases 1, 3, and 4 shows that adding the copper to the mask
elements reduces the detected lead fluorescence so that it is consistent with
the no-mask case (case 2).   The relatively
weak lead-fluorescence detection through the side detector surfaces
(case 8) is due to the low probability that a sufficiently energetic photon
will be photoelectrically absorbed in the side-shield lead
and result in the emission of a fluorescence photon which subsequently enters
the detector through one of the side surfaces; most of these
lead-fluorescence photons would have a higher probability of entering the
detector through the front or rear surfaces, because of the larger solid
angles they subtend as viewed from most points on the side shields.
The 75 keV line flux which is detected in the flight configuration
(case 1) is due to the lead in the collimator blades and could be
reduced by adding more shielding on top of the lead, at the expense of reducing
the open collimator area.  The tin-fluorescence fluxes shown in
Table 2 indicate that, without copper shielding, there would be significant
contributions from the mask (case 4) and rear shield (case 12), but not
from the side shields (case 9).
The tin-fluorescence flux and upper limits listed for cases 1--6 in Tables 1
and 2 are the same because, for these payload configurations, the only photons
reaching the detector arrive on the front surface.
The 100 keV continuum fluxes shown in Table 1 are all comparable to each other
except for the cases with the collimator (case 5), side shields (case 7),
or rear shield (case 10) removed, where they are significantly higher.

As shown in Figures 6(a--c), the total simulated background spectrum agrees
reasonably well (within a factor of $\sim$ 2, when the $^{241}$Am calibration
source counts are removed) with the measured spectra.
The differences between the simulated and measured spectra can be attributed
to our simplified representation (e.g., neglect of the surrounding gondola
structure and MIXE1 detector) of the payload and other approximations such as
using input spectra which were derived from data acquired at a specific
altitude and location, even though the actual coordinates of the EXITE2
payload differed somewhat during its flight.   In addition, a high voltage
related noise problem present during flight (but not while the payload was on
the ground) degraded our PSD
performance, particularly at the lower energies ($<$ $\sim$ 60 keV),
preventing us from making more
restrictive rise-time cuts which would have reduced the residual background
at these energies.  The cause of this problem has since been identified and
fixed.

With the telescope
stowed, the measured background at 100 keV is $\sim$ 4 $\times$ 10$^{-4}$
counts cm$^{-2}$ s$^{-1}$ keV$^{-1}$, which agrees closely with the simulated
results and is slightly lower than the background measured by the Caltech
group [17] for an imaging phoswich detector above Alice Springs,
Australia.  The Caltech detector has a larger field of view than the EXITE2
detector (15$^{\circ}$ vs. 4.7$^{\circ}$, FWHM), making its aperture-flux
background higher.  We note that the difference in geomagnetic latitude between
Alice Springs and Palestine normally yields background rates a factor of
$\sim$ 2 lower in Alice Springs.
With the telescope unstowed, the measured background at 100 keV is $\sim$
2.5 $\times$ 10$^{-4}$ counts cm$^{-2}$ s$^{-1}$ keV$^{-1}$.
The measured background is generally higher with the telescope
stowed due to increased Compton scattering into the field of view from the
gondola structure above the detector, as well as the generally higher
background environment associated with the lower altitude at which the stowed
data were obtained.  The atmospheric 511 keV line
is barely detectable above the continuum and appears as a slight excess in the
measured spectra.  The observed intensity of the line is
consistent with the flux measured by Schindler {\it et al.} [17], given the
smaller field of view of the EXITE2 telescope.  The line is not included in
the simulated results.

The flux sensitivity of a coded-aperture telescope, in the applicable limit
where the detector background counting rate is much larger than the counting
rate due to the astrophysical source, is given by

\begin{equation}
F_{K} = \frac{2K}{\eta\epsilon t_{a} t_{i} t_{c} (1 - t_{m})}{\sqrt{\frac{B}{A \tau \Delta E}}}
\end{equation}

\noindent
where $F_{K}$ is the source flux, in photons
cm$^{-2}$ s$^{-1}$ keV$^{-1}$,
which can be detected with statistical significance $K$; $\eta$ is the imaging
efficiency, a factor ($<$ $\sim$ 1) which accounts for the loss of sensitivity
due to finite detector spatial resolution (at any given energy) relative to the
coded-aperture
mask cell size [18]; $\epsilon$ is the photon full-energy detection efficiency;
$t_{a}$ is the fraction of source photons transmitted through the overlying
atmosphere; $t_{i}$ is the fraction of source photons transmitted through
passive material in the instrument (i.e., the aluminum detector entrance
window and the Lexan coded-aperture support sandwich); $t_{c}$ is the fraction
of source photons transmitted through the collimator and is taken as 0.844
for normal incidence; $t_{m}$ is the fraction of source photons transmitted
through closed coded-aperture cells; $B$ is the detector background counting
rate, in counts cm$^{-2}$ s$^{-1}$ keV$^{-1}$; $A$ is the geometrical
usable area of the detector, in cm$^{2}$; $\tau$ is the observation time in
seconds; and $\Delta$$E$ is the width, in keV, of the energy band for the flux
measurement.

The 3 $\sigma$ detection sensitivity, based on the simulated background
spectrum, is plotted as a function of energy in
Figure 7, assuming 4.0 g cm$^{-2}$ residual atmosphere along the line of sight
(corresponding to 3.5 g cm$^{-2}$ residual atmosphere and a telescope
elevation of $\sim$ 60$^{\circ}$), 10$^{4}$ s observation
time, and an energy bandwidth ${\Delta}E$ equal to 50\% of the energy.
The sensitivity is plotted for NaI(Tl) alone and for NaI(Tl) and CsI(Tl)
combined (i.e., with PSD turned on and off,
respectively).  At 100 keV, PSD rejection is found to reduce the instrument
background by a factor of $\sim$ 3 and to improve the sensitivity by a factor
of $\sim$ 1.5.  At energies
above $\sim$ 215 keV, it is advantageous to retain CsI(Na) events since the
additional background is more than compensated for by the additional stopping
power of the CsI(Na), resulting in improved sensitivity.

We thank T. Gauron, V. Kuosmanen, F. Licata, G. Nystrom, R. Scovel, and
S. Collins for technical support.  This work was partially supported by the
National Aeronautics and Space Administration (NASA) under grant NAGW-624.

\newpage

\clearpage

%\centerline{\bf\large TABLE I}
\footnotesize
\begin{table*}[h]
\begin{center}
\hspace*{-1.5cm}
\begin{tabular}{|l|l|l|l|l|}
\hline
%Case & Payload & F$_{25\rm{keV}}$ & F$_{74\rm{keV}}$ & F$_{100\rm{keV}}$\\
Case & Payload & $F_{25keV}$ & $F_{75keV}$ & $F_{100keV}$\\
& Configuration & (ph cm$^{-2}$ s$^{-1}$) & (ph cm$^{-2}$ s$^{-1}$) & (ph cm$^{-2}$ s$^{-1}$ keV$^{-1}$)\\
\hline
1 & 111 11 111 111 & $<$ 3.70 $\times$ 10$^{-5}$ & (6.44 $\pm$ 1.56) $\times$ 10$^{-4}$ & (1.13 $\pm$ 0.05) $\times$ 10$^{-3}$\\
\hline
2 & 000 11 111 111 & $<$ 5.40 $\times$ 10$^{-5}$ & (7.27 $\pm$ 1.64) $\times$ 10$^{-4}$ & (1.11 $\pm$ 0.05) $\times$ 10$^{-3}$\\
3 & 100 11 111 111 & $<$ 3.33 $\times$ 10$^{-5}$ & (6.40 $\pm$ 1.57) $\times$ 10$^{-4}$ & (1.14 $\pm$ 0.05) $\times$ 10$^{-3}$\\
4 & 110 11 111 111 & (1.31 $\pm$ 0.30) $\times$ 10$^{-4}$ & (7.48 $\pm$ 1.58) $\times$ 10$^{-4}$ & (1.08 $\pm$ 0.05) $\times$ 10$^{-3}$\\
\hline
5 & 111 00 111 111 & $<$ 2.52 $\times$ 10$^{-3}$ & $<$ 2.93 $\times$ 10$^{-3}$ & (8.38 $\pm$ 0.32) $\times$ 10$^{-3}$\\
6 & 111 10 111 111 & $<$ 3.71 $\times$ 10$^{-5}$ & (8.28 $\pm$ 1.63) $\times$ 10$^{-4}$ & (1.16 $\pm$ 0.05) $\times$ 10$^{-3}$\\
\hline
7 & 111 11 000 111 & $<$ 2.66 $\times$ 10$^{-4}$ & (1.26 $\pm$ 0.45) $\times$ 10$^{-3}$ & (3.29 $\pm$ 0.11) $\times$ 10$^{-3}$\\
8 & 111 11 100 111 & $<$ 3.90 $\times$ 10$^{-5}$ & (8.61 $\pm$ 1.73) $\times$ 10$^{-4}$ & (1.13 $\pm$ 0.05) $\times$ 10$^{-3}$\\
9 & 111 11 110 111 & $<$ 4.24 $\times$ 10$^{-5}$ & (5.54 $\pm$ 1.59) $\times$ 10$^{-4}$ & (1.12 $\pm$ 0.05) $\times$ 10$^{-3}$\\
\hline
10 & 111 11 111 000 & $<$ 5.27 $\times$ 10$^{-5}$ & (1.00 $\pm$ 0.33) $\times$ 10$^{-3}$ & (3.12 $\pm$ 0.10) $\times$ 10$^{-3}$\\
11 & 111 11 111 100 & $<$ 3.52 $\times$ 10$^{-5}$ & (4.28 $\pm$ 0.21) $\times$ 10$^{-3}$ & (1.08 $\pm$ 0.05) $\times$ 10$^{-3}$\\
12 & 111 11 111 110 & (1.66 $\pm$ 0.09) $\times$ 10$^{-3}$ & (6.01 $\pm$ 1.55) $\times$ 10$^{-4}$ & (1.13 $\pm$ 0.05) $\times$ 10$^{-3}$\\
\hline
\end{tabular}
\caption{Simulated photon-induced background.  For each shielding
configuration,
the table lists which payload components (mask lead, tin, and copper;
collimator lead and copper; side-shield lead, tin, and copper; rear-shield
lead, tin, and copper) are present (indicated by a ``1'' in the corresponding
column; for example, case 3
has the mask lead present, mask tin and copper both absent, and all other
components present), and the tin-fluorescence line flux or 2 $\sigma$ upper
limit (in photons cm$^{-2}$ s$^{-1}$, with the $K_{\alpha_1}$ and
$K_{\alpha_2}$ lines at $\sim$ 25 keV combined) incident on the detector,
lead-fluorescence line flux or 2 $\sigma$ upper limit
($K_{\alpha_1}$ line at $\sim$ 75 keV), and continuum flux
(in photons cm$^{-2}$ s$^{-1}$ keV$^{-1}$) at 100 keV, with 1 $\sigma$
uncertainties also listed.  Uncertainties are based on Poisson noise
considerations and could have been reduced by running longer simulations.}
\end{center}
\end{table*}

\newpage
\clearpage

%\centerline{\bf\large TABLE II}
\footnotesize
\begin{table*}[h]
\begin{center}
\hspace*{-1.5cm}
\begin{tabular}{|l|l|l|l|l|}
\hline
Case & Payload & $F_{25keV}$ & $F_{75keV}$ & $F_{100keV}$\\
& Configuration & (ph cm$^{-2}$ s$^{-1}$) & (ph cm$^{-2}$ s$^{-1}$) & (ph cm$^{-2}$ s$^{-1}$ keV$^{-1}$)\\
\hline
1 & 111 11 111 111 & $<$ 3.70 $\times$ 10$^{-5}$ & (3.72 $\pm$ 0.49) $\times$ 10$^{-4}$ & (4.39 $\pm$ 0.92) $\times$ 10$^{-5}$\\
\hline
2 & 000 11 111 111 & $<$ 5.40 $\times$ 10$^{-5}$ & (3.70 $\pm$ 0.56) $\times$ 10$^{-4}$ & (3.67 $\pm$ 0.84) $\times$ 10$^{-5}$\\
3 & 100 11 111 111 & $<$ 3.33 $\times$ 10$^{-5}$ & (5.26 $\pm$ 0.58) $\times$ 10$^{-4}$ & (4.23 $\pm$ 0.90) $\times$ 10$^{-5}$\\
4 & 110 11 111 111 & (1.31 $\pm$ 0.30) $\times$ 10$^{-4}$ & (4.48 $\pm$ 0.53) $\times$ 10$^{-4}$ & (4.42 $\pm$ 0.92) $\times$ 10$^{-5}$\\
\hline
5 & 111 00 111 111 & $<$ 2.52 $\times$ 10$^{-3}$ & $<$ 2.81 $\times$ 10$^{-3}$ & (7.34 $\pm$ 0.31) $\times$ 10$^{-3}$\\
6 & 111 10 111 111 & $<$ 3.71 $\times$ 10$^{-5}$ & (6.07 $\pm$ 0.63) $\times$ 10$^{-4}$ & (3.16 $\pm$ 0.79) $\times$ 10$^{-5}$\\
\hline
7 & 111 11 000 111 & $<$ 8.52 $\times$ 10$^{-6}$ & $<$ 3.39 $\times$ 10$^{-4}$ & (5.18 $\pm$ 0.44) $\times$ 10$^{-4}$\\
8 & 111 11 100 111 & $<$ 4.62 $\times$ 10$^{-6}$ & (1.80 $\pm$ 0.60) $\times$ 10$^{-4}$ & (1.04 $\pm$ 0.15) $\times$ 10$^{-4}$\\
9 & 111 11 110 111 & $<$ 4.45 $\times$ 10$^{-6}$ & $<$ 1.03 $\times$ 10$^{-4}$ & (8.70 $\pm$ 1.30) $\times$ 10$^{-5}$\\
\hline
10 & 111 11 111 000 & $<$ 7.85 $\times$ 10$^{-6}$ & $<$ 6.20 $\times$ 10$^{-4}$ & (2.95 $\pm$ 0.10) $\times$ 10$^{-3}$\\
11 & 111 11 111 100 & $<$ 4.57 $\times$ 10$^{-6}$ & (3.87 $\pm$ 0.19) $\times$ 10$^{-3}$ & (9.26 $\pm$ 0.43) $\times$ 10$^{-4}$\\
12 & 111 11 111 110 & (1.63 $\pm$ 0.09) $\times$ 10$^{-3}$ & $<$ 2.74 $\times$ 10$^{-4}$ & (9.64 $\pm$ 0.43) $\times$ 10$^{-4}$\\
\hline
\end{tabular}

\caption{Simulated photon-induced background.  Same as Table 1, except that
here fluxes are measured for photons incident only on the front (cases 1--6),
side (cases 7--9), and rear (cases 10--12) detector surfaces.  For cases 1--6,
the tin-fluorescence flux and upper limits are the same as in Table 1, since
at this energy the only photons reaching the detector arrive on the front
surface, for the corresponding payload configurations.}
\end{center}
\end{table*}

\end{document}